\documentclass[twocolumn,showpacs,amsmath,amssymb, showkeys]{revtex4}

\usepackage{graphicx}
\usepackage{dcolumn}
\usepackage{bm}

\begin{document}
\openup6pt 

\title{Compatibility of Einstein minimally coupled self interacting scalar field theory with the solar system tests of gravity}

\author{A. Bhadra} 
\email{aru_bhadra@yahoo.com}
\affiliation{High Energy $\&$ Cosmic Ray Research Centre, University of North Bengal, Siliguri, WB 734013 
India}

\begin{abstract}
We examine the compatibility of the Einstein minimally coupled self-interacting scalar field theory with the local tests of gravity. We find that apart from the trivial case of the Schwarzschild-de Sitter solution with constant scalar field the theory does not admit any other {\it static} solution, which is consistent with the solar system tests of gravity. 
\end{abstract}

\pacs{98.80.Cq,  04.80.Cc,  95.36.+x}    
\keywords{Scalar field with potential, weak field tests}
\maketitle 

\section{introduction}
A fundamental long range global scalar field $\varphi$ with a potential $V(\varphi)$ is often invoked in different context,  predominantly in cosmology, in order to surmount any shortcomings of the standard theory of gravitation. For instance, to overcome the {\it horizon} and {\it flatness} problems of the standard {\it big bang} model, such a scheme is usually employed to achieve the inflationary phase of the universe [1]. In other areas of physics also, scalar field is found ample applications. For instance, masses of standard model particle are also based on scalar fields with non-vanishing potential, the so called Higgs field. The justification of introducing scalar fields in gravity sector comes from the fact that they arise generically [2], in addition to the usual tensor fields, in most efforts of unifying gravity with all other fundamental interactions, such as the superstring theory or modern revival of Kaluza-Klein theory.  

The inclusion of a pervading scalar field in gravity sector is, however, problematic in one aspect - exterior of a spherically symmetric black hole does not admit any nontrivial classical scalar field owing to the {\it no scalar hair} theorems [3,4]. Such a feature implies that the exact non-trivial solutions of Einstein scalar field theory must possess naked (visible) singularities. On the other hand the well-known {\it cosmic censorship} conjecture [5,6] forbids development of naked singularity generically in realistic gravitational collapse. However, proofs of {\it no scalar hair} theorems [3] rely on positivity of scalar potential ({\it i.e.} on satisfying of weak energy conditions), particular asymptotic conditions and symmetries of spacetime. Many of such restrictions have little justifications in view of recent development in cosmology [7] and string theory [8]. By relaxing the conditions of asymptotic flatness and/or positivity of potential, in recent years several scalar hairy black holes have been advanced in the literatures for some definite scalar potentials [9] though the positivity of total energy seems violated in those cases [4]. Nevertheless, intensive efforts over the years of disproving the no scalar hair conjecture by discovering scalar hairy black holes indicate that the conjecture has not yet been established convincingly for all realistic situations. The naked singularity solutions are also not yet ruled out as the question of cosmic censorship is still open [5] due to non-availability of any rigorous proof of the conjecture. Some investigations even claim that there are classes of collapse evolutions those lead to formation of naked singularities for some given reasonable initial density and pressure profiles of a matter cloud [10]. 

In the presence of global ubiquitous minimally coupled scalar field obviously the components of stress energy tensor do not vanish for vacuum (matter free state) but are equal to those of the background scalar field. Consequently the solutions of the Einstein minimally coupled self-interacting scalar field (EMCSISF) theory under matter free condition describe exterior gravitational field due to a gravitating object instead of vacuum solutions of general relativity. Now if fundamental self-interacting scalar field does exist in gravity sector the static spherically symmetric matter free solution(s) of the EMCSISF theory must produce standard solar system tests non-trivially in the weak field limit irrespective of the black hole or naked singularity nature of the solution. The Buchdahl-JNW solution [11] of the theory for vanishing scalar potential is known to satisfy such a condition but so far the question of consistencies of the EMCSISF theory with solar system or other local tests for non vanishing scalar potential has not received much attention. This is probably because of the common perception that scalar potential can be ignored locally as the mass of scalar field is expected to be small enough. Note that the basic theory (general relativity) remains unaltered for minimal coupling of scalar field.  

In this work we have shown that contrary to the conventional wisdom the EMCSISF theory is incompatible with the solar system tests unless solar gravity is dynamic. In other words whatever be the choice of scalar potential, other than the trivial Schwarzschild-de Sitter solution with constant scalar field the theory does not admit any other {\it static} solution that can explain local observations.   

\section{The spacetime geometry of the solar system}

The general static, spherically symmetric metric in the isotropic coordinates is given by  

\begin{equation}
ds^{2}= -B(\rho) dt^{2} + A(\rho) \left[d\rho^{2} +\rho^{2}\left (d\theta ^{2} + sin^{2} \theta d\phi ^{2} \right) \right] \;,
\end{equation}

For compatibility with the solar field tests of gravity the metric coefficients of Eq.(1) must have the form [12]

\begin{equation}
B = 1- \frac{2m}{\rho} + 2\beta \frac{m^{2}}{\rho^{2}} + {\cal O} \left(\frac{m^{3}}{\rho^{3}} \right) +..
\end{equation}

and

\begin{equation}
A=1+\frac{2\gamma m}{\rho}-\frac{3\delta m^{2}}{2\rho^{2}} + {\cal O} \left(\frac{m^{3}}{\rho^{3}} \right) +.. 
\end{equation}

where $\gamma $ and $\beta$ are the post Newtonian (PN) parameters, [13,14] and $\delta $ is a second-PN parameter [15]. Solar system experiments currently set the bounds $\gamma  = 1+ (2.1 \pm 2.3) \times 10^{-5}$ and $\beta < 1 + 6 \times 10^{-4}$ [12]. The second-PN parameter $\delta$ is so far unconstrained by the solar system tests (however, see [16]). In general relativity, all of these PN parameters are equal to $1$ whereas in minimally coupled scalar field theory with vanishing scalar potential, though both $\gamma$ and $\beta$ are equal to $1$ but $\delta$, which is a measure of scalar charge, is totally arbitary [17]. Note that both $A$ and $B$ may contain higher order terms in $\frac{1}{\rho}$ as well as terms in $\rho$ but the magnitude of such terms must be much smaller than those given explicitly in the Eqs.(3) and (4). The coefficients of the higher order and other terms are unrestricted by the solar system tests conducted so far.

\section{Compatibility of Einstein minimally coupled self interacting scalar field theory with solar system gravity}

Different gravitational effects due to the Sun, such as the gravitational lensing or gravitational time delay, have been detected/measured so far only at external points (in matter free region). For slowly moving bodies and weak interbody gravity the parameterized post-Newtonian (PPN) metric describes gravitational field at external points due to a reasonable matter distribution [13,14] for a broad class of metric theories including general relativity. On the other hand gravitational field at external points also can be obtained from the exterior (outside the matter distribution) solutions of a gravitational theory, like the Schwarzschild solution of general relativity but such solutions contain free parameters which are usually fixed from the boundary conditions (such as by comparing with the parameterized post-Newtonian (PPN) metric in the weak field) and thereby incorporating the effects of the central matter distribution. 
     
We consider the matter free action for a self-gravitating scalar field with an arbitrary potential V($\varphi$) (we use geometrized units such that $G=c=1$ and follow the signature -,+,+,+)

\begin{equation}
{\cal A}= \frac{1}{16 \pi }\int d^{4}x \sqrt{-g}\left( R + \epsilon g^{\mu\nu} \varphi_{,\mu} \varphi_{,\nu} -2V(\varphi) \right) 
\end{equation}

where $R$ is the Ricci scalar, $\epsilon$ equals to $+1$ and $-1$ correspond to normal and ghost (negative scalar kinetic energy) scalar field respectively. 

For solar system gravity, the scalar field must have spherical symmetry as demanded by the field equations but it may not be independent of $t$ in general (though in most non-cosmological cases involving scalar field it is conjectured as static). The field equation $G_{t\rho}= \kappa T_{t\rho}$, however, suggests 

\begin{equation}
\varphi' \dot{\varphi} = 0,
\end{equation}

where the prime and dot, respectively, represent partial differentiation with respect to $\rho$ and $t$. The Eq.(5) immediately implies that either $\varphi' =0$ or $\dot{\varphi} =0$ (excluding the trivial constant $\phi$ solution). First we would consider the case of static scalar field i.e. when $\varphi\equiv \varphi(\rho)$. The field equations accordingly read

\begin{equation}
\left( \frac{A'}{A}\right)^{2} + \frac{1}{2}\frac{A'}{A}\frac{B'}{B} - \frac{1}{\rho} \left( \frac{A'}{A} -\frac{B'}{B} \right) -\frac{A''}{A}  = \epsilon \varphi'^{2}
\end{equation}

\begin{equation}
\frac{1}{A}\left[ \frac{1}{4}\frac{A'}{A}\frac{B'}{B} + \frac{1}{\rho} \frac{B'}{B}  -\left( \frac{B'}{2B}\right)^{2}+ \frac{B''}{2B}  \right] =  - V(\varphi) 
\end{equation}

\begin{equation}
\frac{A''}{2A}- \left( \frac{A'}{2A}\right)^{2}+ \frac{3}{2\rho}  \frac{A'}{A}- \frac{B''}{2B}+ \left( \frac{B'}{2B}\right)^{2}- \frac{1}{2\rho}  \frac{B'}{B} = 0
\end{equation}

(the proportionality constant $\kappa$ has been absorbed in scalar field/potential). The Klein-Gordon equation takes the form

\begin{equation}
\frac{1}{A \rho^{2} \sqrt{AB} }  \left( \rho^{2}\sqrt{AB} \varphi' \right)^{'} = \epsilon \frac{dV}{d\varphi}
\end{equation}

Note that only three out of these four equations (6)-(9) are independent. Before proceeding further here we would like to reveal our strategy first. When the metric coefficients $A$ and $B$ are known, the scalar field and the potential are completely fixed by the Eqs.(6) -(9). However, the weak field tests do not specify the metric coefficients $A$ and $B$ {\it exactly}; the local observations allow a small deviation of first and second PN parameters from the GR value and impose practically no restriction on the higher order PN parameters. Consequently the scalar field and potential should enjoy some flexibility. Our target is to identify the range of scalar potentials those are admissible by the weak field tests. 

Let us first consider the simplest case of constant $V$ i.e. when the potential $V$ is independent of $\varphi$. In such a situation the Eq.(9) implies that either $\varphi$ is a constant and thereby reduces to the cosmological constant scenario, or to the leading order $\varphi \sim 1/\rho$. However, the later solution of scalar field (along with the constant V) does not simultaneously satisfy the Eqs.(6) and (7) unless $V$ vanishes (here our interest is restricted to only self interacting scalar field theory i.e. when scalar potential is non-zero).
The above conclusion is more revealing from Eqs.(12), (13) and (15) (to be introduced later). 

For a non-trivial $V(\varphi)$ the Eq.(6) yields 

\begin{equation}
\sqrt{\epsilon} \varphi = \varphi_{\infty} \pm  \xi \sqrt{1-\gamma} \pm \xi^{2} \sqrt{4\beta +6 \delta-10 } + {\cal O} \left(\xi^{3}\right) +..
\end{equation}
    
where $\varphi_{\infty}$ is the asymptotic (constant) value of scalar field and $\xi = \left(\frac{m}{\rho} \right)^{1/2}$. For the solar system $\xi$ is at most $1.5 \times 10^{-3}$. Substituting  $\varphi$ from the Eq.(10) to the Eq.(9) and considering only upto the first order term in $\frac{m}{\rho}$, one obtains $V(\varphi) \sim \left(\varphi - \varphi_{o} \right)^{6}$. Under the same consideration the Eq.(7), however, implies that $V(\varphi) \sim \left(\varphi - \varphi_{o} \right)^{8}$. Such an inconsistency disappears only when $\gamma =1$, which means that whatever be the choice of potential the EMCSISF theory is indistinguishable from GR at the first post-linear level. 

Next we assume that the effect of scalar field occurs at the next higher order terms in $\frac{m}{\rho}$ so that $\beta$ and/or $\delta$ are different from the canonical GR value of $1$. But when the terms involving $\beta$ and $\delta$ are taken into account in the expressions for metric tensors and scalar field with $\gamma =1$ the same discrepancy regarding $V$ as stated in the above paragraph again crops up {\it i.e.} the Eqs. (7) and (9) do not yield any consistent $V(\varphi)$ under such consideration. To overcome it $4\beta+6\delta$ must be set to $10$. Since $\beta$ is experimentally constrained to nearly $1$, the stated condition demands that $\delta$ has to be also close to $1$ for the EMCSISF theory, but it does not require $\beta$ and $\delta$ to take their exact GR value. A small deviation from GR thus seems admissible. However, to conclude that the EMCSISF theory admits a non-trivial (different from that of general relativity) and solar system compatible solution, it needs to show that in the weak field limit the theory does produce spacetime metric of the form as given by the Eqs.(2) and (3) with a non-trivial scalar field. 

At this stage for convenience we would switch over to a new coordinate system in which the general static, spherically symmetric metric takes the form  

\begin{equation}
ds^{2}= -f(r) dt^{2} + f^{-1}(r) dr^{2} +R^{2}(r)\left (d\theta ^{2} + sin^{2} \theta d\phi ^{2} \right)
\end{equation}

and accordingly $\varphi=\varphi(r)$. The field equations in this coordinate system read

\begin{equation}
\left(f' R^{2}\right)^{'} = -2 R^{2} V(\varphi)
\end{equation}

\begin{equation}
2 R^{''}/R = -\epsilon \varphi'^{2}
\end{equation}

\begin{equation}
f(R^{2})^{''} - R^{2}f^{''}=2
\end{equation}

and the wave equation is given by

\begin{equation}
(f R^{2}\varphi')^{'} = \epsilon R^{2}\frac{dV}{d\varphi}
\end{equation}

where the prime denotes $d/dr$. To the leading order the transformation equation that relates this new coordinate system to the isotropic coordinate system with $\gamma=1$ is just $r \sim r_{o}+\rho$, $r_{o}$ being a constant. The GR expressions of the metric coefficients in this new coordinate system are $f = 1-2m/r $ and $R = r$. 

Suppose due to the scalar field effect, $f$ and $R$ have been modified and thus contain higher order terms in $m/r$ and the scalar field takes the following general form

\begin{equation}
\sqrt{\epsilon} \varphi = \varphi_{o} + \frac{a_{n}}{ r^{n/2}} + \mbox{higher order terms in 1/r} 
\end{equation}

where $a_{n}$ is the coefficient of the first non vanishing term in the power series of $1/r$, the index $n$ is a positive integer and is greater than $2$. From the Eq.(13) one obtains that to the leading order $R \sim r \left( 1-\frac{b}{r^{n}} \right) $, where $b=\frac{na_{n}^2}{8(n-1)}$. Consequently the Eq.(15) yields that $V(\varphi) \sim \left(\varphi-\varphi_{o}\right)^{2+4/n}$ whereas the Eq.(12) implies that $V(\varphi) \sim \left(\varphi - \varphi_{o}\right)^{8/n}$. Thus contradiction remains. The power index n can be negative as well. But the stated inconsistency persists for negative $n$ also unless $a_{n}$ vanishes. 

Next we consider the case of non-static scalar field. In that case $\varphi'=0$ i.e. scalar field is independent of space coordinates. The field equations for the general metric (1) are then given by
 
\begin{equation}
\frac{B}{A} \left[ \frac{A'^{2}}{A^{2}} + \frac{A'B'}{2AB} - \frac{1}{\rho} \left( \frac{A'}{A} -\frac{B'}{B} \right) -\frac{A''}{A} \right] = \epsilon \dot{\varphi}^{2}
\end{equation}

\begin{equation}
\frac{1}{A}\left[ \frac{A''}{2A} + \frac{B''}{2B} - \frac{A'B'}{4AB} - \frac{A'^{2}}{A^{2}} - \frac{B'^{2}}{4B^{2}}+ \frac{A'}{A \rho}  \right] =  - V(\varphi) 
\end{equation}

\begin{equation}
\frac{B}{A} \left( \frac{B''}{2B} - \frac{B'^{2}}{4B^{2}} + \frac{B'}{2 B \rho}- \frac{A''}{2A} +  \frac{A'^{2}}{4A^{2}} -  \frac{3A'}{2 A \rho}  +  \right) = \epsilon \dot{\varphi}^{2}
\end{equation}

The Klein-Gordon equation reads $\ddot{ \varphi} = \epsilon \frac{dV}{d\varphi}$. The left hind side of the Eqs.(17)-(19) contain terms involving only $\rho$. Since $\varphi$ is independent of $\rho$, the coefficients of all the powers of $\rho$ must vanish separately and $\dot{\varphi}$ only can be a constant (hence at best $\varphi$ can be a linear function of $t$). The wave equation immediately implies that $V$ is also a constant, independent of scalar field and hence there is no self interaction. Further it can be readily checked that the PPN metric (Eqs. (1) -(3)) (with added terms of different powers in $\rho$) does not consistently satisfy the above field equations.  

\section{Conclusion}

It is clear from the discussions of the previous section that the EMCSISF theory does not admit any {\it static} solution which is consistent with the local tests of gravity except the trivial case of the Schwarzschild-de Sitter solution with constant scalar field. Several exact/numerical solutions of the EMCSISF theory for different (particular) scalar potential exist in the literature [9]. All of those solutions are found in accordance with the present findings as expected. If a massive scalar field exists in the gravity sector then it should produce some observational effects, which even may not be detectable with present technology. In a metric theory the effects of gravity are revealed through the space-time metric. The Schwarzschild-de Sitter metric, which is the solitary solar system compatible solution of the EMCSISF theory, doest not bear any signature of scalar field and hence there will be no effect of scalar field whatsoever on a test particle (this is natural because for constant scalar field, the EMCSISF theory reduces to (pure) general relativity with a cosmological constant. The present finding thus raises serious doubt on the existence of massive scalar field in gravity sector.    
In the present work our discussion was limited to the case of minimal coupling of scalar field to space-time curvature. Scalar field also can be coupled to gravity nonminimally (as well as conformally) so that gravity sector itself is modified [18]. In that case the theory enjoys an additional degree of freedom in terms of scalar field coupling function and its predictions for physical effects can differ from those of the EMCSISF theory substantially [17,18,19]. Still it seems worthwhile to examine the consistencies of the scalar tensor theory (generalized Jordan-Brans-Dicke theory) with arbitrary potential with the weak field tests following the idea of the present work. 

\section*{Acknowledgments}

The author would like to thank two anonymous referees for insightful comments and suggestions.

\end{document}